\documentstyle[aps,preprint,pra]{revtex}
\begin{document}
\newcommand{\etal}{{\em et al.}\/}
\newcommand{\IP}{inner polarization}
\newcommand{\IPF}{\IP\ function}
\newcommand{\IPFs}{\IP\ functions}
\newcommand{\auth}[2]{#1 #2, }
\newcommand{\jcite}[4]{#1 {\bf #2}, #3 (#4)}
\newcommand{\et}{ and }
\newcommand{\twoauth}[4]{#1 #2 and #3 #4,}
\newcommand{\oneauth}[2]{#1 #2,}
\newcommand{\andauth}[2]{and #1 #2, }
\newcommand{\book}[4]{{\it #1} (#2, #3, #4)}
\newcommand{\erratum}[3]{\jcite{erratum}{#1}{#2}{#3}}
\newcommand{\inbook}[5]{In {\it #1}; #2; #3: #4, #5}
\newcommand{\JCP}[3]{\jcite{J. Chem. Phys.}{#1}{#2}{#3}}
\newcommand{\jms}[3]{\jcite{J. Mol. Spectrosc.}{#1}{#2}{#3}}
\newcommand{\jmsp}[3]{\jcite{J. Mol. Spectrosc.}{#1}{#2}{#3}}
\newcommand{\ang}[3]{\jcite{Angew. Chem. Int. Ed. Engl.}{#1}{#2}{#3}}
\newcommand{\jmstr}[3]{\jcite{J. Mol. Struct.}{#1}{#2}{#3}}
\newcommand{\joc}[3]{\jcite{J. Org. Chem.}{#1}{#2}{#3}}
\newcommand{\cpl}[3]{\jcite{Chem. Phys. Lett.}{#1}{#2}{#3}}
\newcommand{\cp}[3]{\jcite{Chem. Phys.}{#1}{#2}{#3}}
\newcommand{\pr}[3]{\jcite{Phys. Rev.}{#1}{#2}{#3}}
\newcommand{\jpc}[3]{\jcite{J. Phys. Chem.}{#1}{#2}{#3}}
\newcommand{\jpcA}[3]{\jcite{J. Phys. Chem. A}{#1}{#2}{#3}}
\newcommand{\jpca}[3]{\jcite{J. Phys. Chem. A}{#1}{#2}{#3}}
\newcommand{\jpcB}[3]{\jcite{J. Phys. Chem. B}{#1}{#2}{#3}}
\newcommand{\jcc}[3]{\jcite{J. Comput. Chem.}{#1}{#2}{#3}}
\newcommand{\molphys}[3]{\jcite{Mol. Phys.}{#1}{#2}{#3}}
\newcommand{\cpc}[3]{\jcite{Comput. Phys. Commun.}{#1}{#2}{#3}}
\newcommand{\jcsfii}[3]{\jcite{J. Chem. Soc. Faraday Trans. II}{#1}{#2}{#3}}
\newcommand{\prsa}[3]{\jcite{Proc. Royal Soc. A}{#1}{#2}{#3}}
\newcommand{\jacs}[3]{\jcite{J. Am. Chem. Soc.}{#1}{#2}{#3}}
\newcommand{\ijqcs}[3]{\jcite{Int. J. Quantum Chem. Symp.}{#1}{#2}{#3}}
\newcommand{\ijqc}[3]{\jcite{Int. J. Quantum Chem.}{#1}{#2}{#3}}
\newcommand{\spa}[3]{\jcite{Spectrochim. Acta A}{#1}{#2}{#3}}
\newcommand{\tca}[3]{\jcite{Theor. Chem. Acc.}{#1}{#2}{#3}}
\newcommand{\tcaold}[3]{\jcite{Theor. Chim. Acta}{#1}{#2}{#3}}
\newcommand{\jpcrd}[3]{\jcite{J. Phys. Chem. Ref. Data}{#1}{#2}{#3}}
\newcommand{\hof}[0]{$\Delta$H$_f$$^{298}$}

\draft
\title{
The heats of formation of the haloacetylenes
XCCY [X, Y = H, F, Cl]: basis set limit ab initio results and
thermochemical analysis}
\author{Srinivasan Parthiban and Jan M.L. Martin}
\address{Department of Organic Chemistry,
Kimmelman Building, Room 262,
Weizmann Institute of Science,
IL-76100 Re\d{h}ovot, Israel. {\rm E-mail:} {\tt comartin@wicc.weizmann.ac.il}
}
\author{Joel F. Liebman}
\address{Department of Chemistry and Biochemistry, 
University of Maryland, Baltimore County,
Baltimore, MD 21250, USA
}
\date{{\em Molecular Physics} MP 1.05/319 (Davidson Festschrift); submitted April 22, 2001; Accepted July 16, 2001}
\maketitle
\begin{abstract}
The heats of formation of haloacetylenes are evaluated using the recent 
W1 and W2 ab initio computational thermochemistry methods. These 
calculations involve CCSD and CCSD(T) coupled cluster methods, basis 
sets of up to spdfgh quality, extrapolations to the one-particle 
basis set limit, and contributions of inner-shell correlation, scalar 
relativistic effects, and (where relevant) first-order spin-orbit coupling. 
The heats of formation determined using W2 theory are: 
\hof(HCCH) = 54.48 kcal/mol, 
\hof(HCCF) = 25.15 kcal/mol, \hof(FCCF) = 1.38 kcal/mol, 
\hof(HCCCl) = 54.83 kcal/mol,
\hof(ClCCCl) = 56.21 kcal/mol, and \hof(FCCCl) = 28.47 kcal/mol.
Enthalpies of hydrogenation and destabilization energies relative 
to acetylene were obtained at the W1 level of theory. So doing we 
find the following destabilization order for acetylenes: 
FCCF $>$ ClCCF $>$ HCCF $>$ ClCCCl $>$ HCCCl $>$ HCCH.
By a combination of W1 theory and isodesmic reactions, we show that
the generally accepted heat of formation of 1,2-dichloroethane should be
revised to -31.8$\pm$0.6 kcal/mol, in excellent agreement with a very recent
critically evaluated review.
The performance of compound thermochemistry schemes such as G2, G3, G3X and
CBS-QB3 theories has been analyzed. 

\end{abstract}
\newpage
\section{Introduction}
There is continuing interest in the thermochemistry of 
halogenated species\cite{JL1}, as there is in the reaction 
chemistry of halogenated acetylenes\cite{JL2}. However, there 
remains no direct determination of the heat of formation of any 
haloacetylene --- these species are simply too unstable and/or 
reactive to allow for such measurements\cite{JL3}. Analysis of 
mixed phase equilibration measurements\cite{JL4} result in a 
suggested value of 5.0$\pm$5.0 kcal/mol for difluoroacetylene, 
while the same source gives estimated values of 47.9$\pm$10.0 kcal/mol 
for dichloroacetylene, and 30.0$\pm$15.1 and 51.1$\pm$10.0 kcal/mol
for HCCF and HCCCl, respectively. The error bars are large; the 
estimation methods assume constancy and additivity approximations that 
are inadequately affirmed for the energetics of halogenated 
species\cite{JL1}. In such circumstances, the only means of
acquiring the needed data is through computations.

A number of theoretical studies on thermochemical properties of
halocarbons have been carried out (see Ref. \cite{bozzelli99,berry98}
and references therein) in the past, because of their adverse
effects on the ozone layer and high global warming potentials.
While there have been numerous studies on saturated halocarbons, 
there is a paucity of thermochemical data for haloacetylenes and
the available data are inconsistent. For example, Colegrove and 
Thompson\cite{colegrove97} evaluated the heats of formation of 
several chlorinated hydrocarbons in which they reported the 
$\Delta$H$_f$$^{298}$  of HCCH as 56.47 and 56.24 kcal/mol at the 
G1 and G2 theories respectively, while Rodriquez \etal\cite{rodriquez93} 
found the enthalpies of formation at 298 K for HCCH and HCCCl at the 
MP4STDQ/6-311G(2df,p) level as 51.7 and 51.4 kcal/mol respectively. 
At MP4STDQ/6-311++G(3df,3pd) level, they found the 
$\Delta$H$_f$$^{298}$ of HCCH as 55.7 kcal/mol.  Wiberg and 
Rablen\cite{wiberg93} reported the heat of formation at 0 K of HCCH
as 56.0 kcal/mol at G2 level of theory. It is worth mentioning that
Schaefer and co-workers\cite{schaefer93} investigated the
fluorovinylidene-fluoroacetylene isomerization reaction on
the C$_2$HF singlet ground-state potential energy surface at
the CCSD(T)/TZ2P level of theory. In the study of vinyl chloride
and bromide, Radom and co-workers\cite{radom93} reported the structure 
of chloroacetylene at the CISD(Q) and MP4 level of theory using 
the 6-31G(d) basis set. Very recently, Breidung and Thiel\cite{thiel01}
studied the equilibirium structure and spectroscopic constants of
difluorovinylidene (F$_2$CC) at the CCSD(T)/aug-cc-pVQZ level of theory
including core correlation effects. In this study, they have also 
reported the thermodynamic stability of F$_2$CC (less stable by 
29 kcal/mol) relative to FCCF as well as the barrier height 
(38 kcal/mol) for the isomerization from F$_2$CC to FCCF. 
Earlier, experimental evidence for the existence of F$_2$CC was 
shown by Breidung \etal\cite{thiel97} in the matrix isolation study, 
in which F$_2$CC was generated by 193-nm laser photolysis of FCCF.

Recently, one of us proposed two new computational thermochemistry 
protocols known as W1 and W2 (Weizmann-1 and -2) theory,\cite{w1,w1review} which 
permits the computation of total atomization energies of small molecules 
in the kJ/mol accuracy range. For the total atomization energies of 
its "training set" of molecules, the more economical W1 theory achieved 
a mean absolute error of 0.37 kcal/mol, which goes down to 0.22 kcal/mol 
for the more rigorous W2 theory, for which imperfections in the CCSD(T) 
electron correlation method\cite{Rag89} are the main accuracy-limiting 
factor.  In a subsequent validation study\cite{w1w2validate}, we have shown 
that W1 and W2 theories yield thermochemical data in the kJ/mol accuracy range 
for most of the G2/97 data set that are well described by a single reference
configuration. 
The XCCY [X=H, F, Cl] compounds are well within the applicability range of 
W2 theory even with moderately powerful computing equipment. 

In the present study, we will carry out W2 calculations for all species 
concerned. W1 results will however also be presented, in order to
establish convergence of our calculated results in terms of the level
of theory. In addition, W1 results are used to calculate the enthalpy
of hydrogenation to form substituted alkanes. Finally, we will 
evaluate the performance of some popular computational thermochemistry 
protocols compared to the present benchmark results.

\section{Methods}

Most ab initio calculations (specifically, the CCSD\cite{Pur82} and CCSD(T) 
coupled cluster calculations involved in W$n$ theory) were carried out 
using MOLPRO 98.1\cite{molpro} running on SGI Origin 2000 and Compaq ES40 
minisupercomputers at the Weizmann Institute of Science. 
(For the open-shell calculations on the constituent atoms, the definition 
of the open-shell CCSD and CCSD(T) energies in Ref.\cite{Wat93}, as 
implemented\cite{Ham92}, was employed.) 

Some frequency calculations using the B3LYP (Becke 3-parameter-Lee-Yang-Parr)
density functional method\cite{Bec93,Lee88} involved in the determination
of the zero-point vibrational energies were carried out using Gaussian 98\cite{g98}
running on these same machines. Finally, for comparison purposes, some calculations
using the G2\cite{g2} and G3\cite{g3} theories of Pople and coworkers, as well as
using the CBS-QB3 method of Petersson and coworkers\cite{cbs-qb3}, were carried out
using the implementations of these methods in Gaussian 98.

The rationale and justification of the W1 and W2 methods are
discussed at length elsewhere\cite{w1,w1review}. In the interest of making 
this paper self-contained, we briefly summarize the computational 
protocols here:
\begin{itemize}
\item the geometry is optimized at the appropriate level of theory, which for
W1 is B3LYP/cc-pVTZ+1 and for W2 is CCSD(T)/cc-pVQZ+1, in which cc-pVTZ and
cc-pVQZ are the Dunning correlation consistent\cite{ccecc} polarized valence triple
and quadruple zeta basis sets\cite{Dun89}, respectively, and the suffix
"+1" refers to the addition of a high-exponent $d$ function on all second-row
atoms to cover inner polarization effects\cite{sio}. Only valence electrons
are correlated in the CCSD(T) calculation;
\item a CCSD calculation with only valence electrons correlated is carried
out for a "big" basis set (see below) containing basis functions of at most
angular momentum $L_{\rm max}$;
\item CCSD(T) calculations with only valence electrons correlated are carried 
out for a "small" (at most $L_{\rm max}-2$) and a 
"medium" (at most $L_{\rm max}-1$) basis set (see below);
\item the SCF component of the atomization energy is extrapolated 
geometrically\cite{Fel92} ($E[L]=E_\infty+A/L^5$) to the "medium", 
and "large" basis set results;
\item the CCSD valence correlation component is extrapolated using the
Schwartz-type expression\cite{Sch63,Hal98} $A+B/L^\alpha$ to the "medium" 
and "large" basis set results. $\alpha$=3.0 for W2 theory, 3.22 for W1 theory;
\item the (T) valence correlation component is extrapolated using the same 
expression, but applied to the "small" and "medium" results only;
\item the inner-shell correlation contribution is calculated as the difference
between valence-only and all-electrons-correlated CCSD(T)/MTsmall binding 
energies, where MTsmall stands for the Martin-Taylor 'small' core correlation
basis set\cite{hf,cc} defined in Ref.\cite{w1};
\item the scalar relativistic contribution is obtained as the expectation
values of the one-electron Darwin and mass-velocity operators\cite{Cow76,Mar83}
from an ACPF (averaged coupled pair functional\cite{Gda88}) calculation
with the MTsmall basis set. All electrons are correlated in this step;
\item except for systems in degenerate states (for which an explicit spin-orbit
calculation is required), the spin-orbit contribution is normally derived
from the atomic fine structures of the constituent atoms.
\end{itemize}
In standard W2 theory, the basis sets employed are Dunning's cc-pV$L$Z on H,
aug-cc-pV$L$Z\cite{Ken92} on B--F, and aug-cc-pV$L$Z+2d1f on Al--Cl, where
the "+2d1f" suffix indicates the addition of two high-exponent $d$ and
one high-exponent $f$ function, the exponents being obtained as geometric
series $2.5^n\alpha$, where $\alpha$ is the highest exponent of that
angular momentum present in the underlying cc-pV$L$Z basis set. For the
"small", "medium", and "large" basis sets, $L$=T, Q, and 5, respectively.
In standard W1 theory, the basis sets employed are as described above
except for $L$=D, T, and Q, respectively, for "small", "medium", and
"large". (For Al--Cl, the "small" basis set is aug-cc-pVDZ+2d.)

Our best calculations in the present work were carried out using a minor 
variation on the W2 protocol of Martin and de Oliveira, which we shall 
denote W2h (h standing for hetero-atom) in this paper. It differs from 
standard W2 theory in that the aug-cc-pV$L$Z basis sets are only used 
for group V, VI, and VII elements, while the regular cc-pV$L$Z basis set 
is used for group III and IV elements.

Zero-point energies were obtained from B3LYP/cc-pVTZ+1 harmonic frequencies
scaled by 0.985 in both the W1 and W2 cases, and thermodynamic functions
computed using the RRHO (rigid rotor-harmonic oscillator) method from the
B3LYP/cc-pVTZ+1 rotational constants and harmonic frequencies.  In the
cases of HCCF, HCCCl, and C$_2$H$_2$, anharmonic zero-point energies are
available from the literature for comparison.

\section{Results and discussion}

Optimized geometries at the B3LYP/cc-pVTZ+1 and CCSD(T)/cc-pVQZ+1
levels of theory are given in Table I. Computed total atomization energies 
and heats of formation for the various species discussed here are given 
in Table II, together with the constituent components (SCF, CCSD valence 
correlation, connected triple excitations, inner-shell correlation, scalar
relativistic effects, spin-orbit coupling). Finally, the reaction
energies (and constituent components) for a number of isodesmic
and isogyric reactions discussed below can be found in Table III.

Comparing the W1 and W2h TAE$_e$ (total atomization energy at the bottom
of the well) suggests that the two treatments are in good agreement with
each other, except for C$_2$F$_2$ where the more rigorous W2h method suggests
a TAE$_e$ value about 0.97 kcal/mol lower than W1. Analysis of the individual
contributions reveals that the difference is evenly split between SCF and 
CCSD valence correlation; we also note that HCCF, CH$_3$F and FCCCl all
reveal about half the W1--W2h difference of C$_2$F$_2$.

The W2 and W2h protocols normally call for anharmonic zero-point vibrational
energies (ZPVE values), e.g. from CCSD(T)/cc-pVTZ (or better) quartic force field
calculations. Accurate ab initio anharmonic force fields are available
for acetylene\cite{c2h2} and HCCF\cite{hccf} from the work
of Martin and coworkers, and for C$_2$F$_2$ and C$_2$Cl$_2$ from the
work of Thiel and coworkers\cite{ThielC2F2,ThielC2Cl2}. From comparing
the W1 and W2h entries for the ZPVE in Table II, we see that the
scaled harmonic B3LYP/cc-pVTZ+1 ZPVEs differ by up to 0.24 kcal/mol (C$_2$H$_2$)
from the more rigorous values. Since this is comparable to the expected accuracy
of W2 and W2h theory themselves, it is worthwhile to substitute the anharmonic
values when at all available. For FCCCl, an anharmonic force field calculation
was published\cite{ThielFCCCl}, but the anharmonicity constants or force field
were not explicitly reported in that paper: hence no zero point energy could
be derived. We have adopted the assumption that the small difference between
the scaled B3LYP and the rigorous values could be interpolated between C$_2$F$_2$
and C$_2$Cl$_2$; a similar approximation was adopted for HCCCl (interpolation 
between C$_2$H$_2$ and C$_2$Cl$_2$).

Let us first consider some of the species involved in the isodesmic reactions.
Computed and WebBook\cite{webbook} heats of formation for 
CH$_4$ are, not surprisingly, in perfect agreement. Our W2h value 
for C$_2$H$_2$ agrees to within overlapping uncertainties with 
the WebBook value.
For CH$_3$F, the WebBook lists two heats of formation:
-56.0 kcal/mol from the JANAF tables\cite{JL4} and -59.0 kcal/mol from Lias et
al.\cite{Lia85}. Our calculations clearly favor the former value.
Likewise, two experimental values are available for CH$_3$Cl: 
-19.59$\pm$0.16 kcal/mol from Fletcher and Pilcher\cite{Fle71}
and -20.53$\pm$0.14 kcal/mol from Lacher et al\cite{Lac56}. Our calculations
agree well with the latter value.

Only rather crude estimates are available for the haloacetylenes themselves
(see Table II): our calculations agree with them to within the former's 
stated error bars, but given the magnitude thereof this is a rather
hollow victory.

Let us consider a particularly simple estimation method and associated
constancy and additivity, namely the assumption of thermoneutrality for
the reaction
\begin{equation}
{\rm HC}{\equiv}{\rm CH} + {\rm XC}{\equiv}{\rm CX} \rightarrow 2 {\rm ~HC}{\equiv}{\rm CX}
\end{equation}

For X=CH$_3$  (using values from Ref.\cite{JL9} for the parent acetylene, 
2-butyne, and propyne of 54.5$\pm$0.2, 34.8$\pm$0.3, and 44.2$\pm$0.2 kcal/mol,
respectively), this reaction is slightly exothermic (by 0.9$\pm$0.5 kcal/mol).
This is comfortably close to thermoneutrality, to the extent that it suggests,
as one would expect for hydrocarbons,
constant triply bonded Benson group increments, C$_t$-(C) and C$_t$-(H).
By contrast, for X=F, we find the reaction to be exothermic by 5.42
kcal/mol at the W2h level(see Table III), compared to only 1.03 kcal/mol for X=Cl.
It is noteworthy that in both cases, the SCF component accounts for essentially
all of the effect, suggesting that a simple explanation must exist in the
structure of the zero-order wave function. We also note that the 
exchange reaction C$_2$F$_2$+C$_2$Cl$_2$$\rightarrow$2~FCCCl is mildly
exothermic, by 0.64 kcal/mol, consisting chiefly of an SCF component which
is exothermic by 1.22 kcal/mol and a valence correlation component which
is {\em endo}thermic by 0.52 kcal/mol. 

Furthermore we note that the reactions
\begin{eqnarray}
{\rm HC}{\equiv}{\rm CH} + {\rm CH}_3{\rm X} \rightarrow {\rm HC}{\equiv}{\rm
CX} + {\rm CH}_4\\
{\rm HC}{\equiv}{\rm CX} + {\rm CH}_3{\rm X} \rightarrow {\rm XC}{\equiv}{\rm
CX} + {\rm CH}_4
\end{eqnarray}
are fairly mildly endothermic for X=Cl (2.43 and 3.46 kcal/mol, respectively,
at 298 K), but pronouncedly endothermic for X=F (9.32 and 14.74 kcal/mol,
respectively). For comparison, W1 results are also presented in Table III;
these are in excellent agreement with W2h values.

Considering the bond distances in Table I, we see that the C$\equiv$C
bond distance in the chloroacetylenes changes only marginally from the
acetylene value; in contrast, the C${\equiv}$C bond in HCCF is compressed
by 0.0060 \AA\ at the CCSD(T)/cc-pVQZ+1 level.
The compression from HCCF to FCCF is still more 
pronounced, namely 0.0103 \AA\ at the same level of theory. 
(A parallel trend can be noted in the harmonic frequency for the CC stretch,
which mounts from 2013 cm$^{-1}$ in acetylene\cite{c2h2} over 2279 cm$^{-1}$
in HCCF to 2527 cm$^{-1}$ in C$_2$F$_2$.) The reason for this probably
lies in the changing importance of ionic contributions of the
type Y-C$\equiv$C$^-$~X$^+$: while for acetylene these play a 
nonnegligible role (after all, acetylene is a very weak acid in aqueous 
solution), such valence bond structures are quite unfavorable for mono-
and particularly for difluoroacetylene.

In all haloacetylenes considered here, the C-X bond 
distances are appreciably shorter than in CH$_{3}$X. This is 
consistent with the existence of resonance effects of the type
\begin{equation}
{\rm H-C}{\equiv}{\rm C-X}     \leftrightarrow {\rm H-C}^{-}={\rm C}={\rm X}^{+}
\end{equation}
as are the Wiberg bond orders\cite{Wiberg} (derived from a natural 
population analysis\cite{npa} on the B3LYP/cc-pVTZ+1 wave function)
in Table IV. (The significant vicinal
Wiberg bond orders BO(XC$'$) found there are perhaps a clearer indication for
this type of resonance than the increased X-C bond orders relative to
CH$_{3}$X.) Such resonance would stabilize the haloacetylenes: on the 
other hand, they would be a great deal less favorable in HCCF than in
HCCCl, and definitely less in FCCF than in ClCCCl. In addition, it 
would appear to be less efficient in dihalides than in monohalides:
\begin{equation}
{\rm Y-C}\equiv {\rm C-X} \leftrightarrow {\rm Y-C}^-={\rm C}={\rm X}^+
\end{equation}
The Wiberg bond orders in Table IV do appear to reflect these trends.

A useful probe of substituent effects on acetylene stability is the 
enthalpy of the hydrogenation to form the related substituted 
alkane \cite{100}:

\begin{equation}
{\rm X}-{\rm C}\equiv{\rm C}-{\rm Y} + 2 {\rm ~H}_2 \rightarrow 
{\rm X}-{\rm CH}_2{\rm CH}_2-{\rm Y}
\end{equation}

Before discussing the fluorinated and chlorinated species of direct
interest in this text, we start with simpler, better known substituents.
For the parent acetylene \cite{200} with X = Y = H, the gas phase (355K)
hydrogenation enthalpy was shown to be -75.1 $\pm$ 0.7 kcal/mol by Conn,
Kistiakowsky and Smith in excellent agreement with the presently computed
(W1) value of -75.5 kcal/mol. By contrast, in hydrocarbon solution (and thus
mimicking the gas phase) Rogers, Dagdagan and Allinger \cite{300} showed
that mono-n alkylacetylenes typically have a hydrogenation enthalpy of ca. -69
and di-n-alkylacetylenes have a value of ca. -64 kcal/mol.  Ignoring
temperature and solvent effects, this 5 kcal/mol stabilization per alkyl
group is consonant with general perceptions regarding both
hyperconjugative and hybridization effects of alkylation. Phenylation is
expected to result in a larger, mostly conjugatively derived,
stabilization. Under the same conditions as these alkylated acetylenes,
Davis, Allinger and Rogers \cite{400} showed that phenylacetylene and
n-alkylphenylacetylenes have an enthalpy of hydrogenation of ca. -62 and
-59.6 $\pm$ 0.5 kcal/mol, respectively, documenting the greater stabilizing
effect of phenyl groups over alkyl.

Not surprisingly, there are no direct measurements of the enthalpy
of hydrogenation of any of the haloacetylenes. In order to obtain
such quantity, accurate enthalpy of formation of haloethanes is
required. But reliable experimental thermochemical data 
for the haloethanes is scarce and typically is derived from 
one or two sources. Therefore, we have calculated the enthalpies
of formation of ethane and the haloethanes at the W1 level of theory: 
the results are presented in Table V. Since the core correlation contributions for
1,2-dichloroethane and 1-fluoro-2-chloroethane are computationally 
very expensive to calculate, we have estimated those values using
the MSFT bond equivalent model proposed by Martin \etal\cite{msft}. 
Table VI shows the W1 reaction energies 
for the hydrogenation reactions together with the constituent components.

We first note that the the W1 calculated \hof of C$_2$H$_6$ 
value (-21.18 kcal/mol) differs by about 1.1 kcal from the experimental 
value\cite{gurvich} of -20.08 $\pm$ 0.10 kcal/mol. 
(Note that the discrepancy 
is reduced to -0.7 kcal/mol at the W2 level, with some of
the remaining error probably due to lack of account for internal rotation
in the zero-point and thermal corrections.)
From Table VI, 
it also be seen that the acetylene hydrogenation energy calculated at the W1 
level (-75.54 kcal/mol) is in excellent agreement 
with the experimental value of -75.51 $\pm$ 0.7 kcal/mol\cite{200}.

In the literature, directly or even indirectly measured enthalpy of
formation of C$_2$H$_5$F are lacking.  But we find few theoretical values
based on empirical schemes, -65.06 kcal/mol from Smith\cite{smith98}
and -66.30 $\pm$ 1.00 kcal/mol from Luo and Benson\cite{luobenson97}.
In addition, there are enthalpy of combustion measurements of some 
long-chain n-alkyl fluorides by R\"uchardt, et al. \cite{600}.  
In particular, for the fluorinated nonane, dodecane and tetradecane, 
enthalpies of formation of -101.2 $\pm$ 0.6, -116.9 $\pm$ 0.2 
and -127.4 $\pm$ 0.2 kcal/mol were reported. These correspond to a 
methylene increment of ca. 5.3 kcal/mol, quite close to the 
4.93 kcal/mol generally recommended\cite{700}. Extrapolating to the 
two carbon ethyl fluoride results in an estimated enthalpy of 
formation of about -65 kcal/mol. In this study, we have calculated 
the $\Delta$H$_f$$^{298}$ of C$_2$H$_5$F as -66.41 kcal/mol at 
the W1 level of theory, favoring the empirical value obtained by 
Luo and Benson. Combining the W1 value with our suggested heat of 
formation of 24.76 kcal/mol for HCCF, we deduce an enthalpy of 
hydrogenation of -91.12 kcal/mol, and so a calculated destabilization 
of 15.58 kcal/mol is found.

Concerning C$_2$H$_5$Cl, we find two experimental values for the 
enthalpy of formation of C$_2$H$_5$Cl, -26.84 $\pm$0.18 kcal/mol 
from Fletcher and Pilcher and -25.72 $\pm$ 0.14 kcal/mol from 
Lacher et al\cite{Lac56}. In the recent W1/W2 validation 
study\cite{w1w2validate} we reported the 
$\Delta$H$_f$$^{298}$[C$_2$H$_5$Cl] = -27.75 kcal/mol at 
the W1 level of theory.  The discrepancy with Fletcher and
Pilcher is nearly identical to the W1 error for the parent
ethane molecule: hence our calculations would appear to
favor this value.
Combining the reported 
W1 value of $\Delta$H$_f$$^{298}$[C$_2$H$_5$Cl] 
with the presently calculated W1 enthalpy of formation of HCCCl 
(54.90 kcal/mol), we derive a hydrogenation enthalpy of 
-82.61 kcal/mol. In other words, HCCCl is destabilized
by ca. 7.07 kcal/mol relative to acetylene.

To our knowledge, no experimental enthalpies of formation
have been reported for 1,2-difluoroethane and 1-fluoro-2-chloroethane.
We have calculated the W1 enthalpy of formation of trans-1,2-difluoroethane 
and trans-1-fluoro-2-chloroethane needed for
evaluating the destabilization of FCCF and FCCCl, respectively. Combining
the values of \hof of FCH$_2$--CH$_2$Cl (-71.06 kcal/mol) and
FC$\equiv$CCl (28.12 kcal/mol) yields hydrogenation and destabilization
energies of -99.14 and 23.60 kcal/mol, respectively. For FCCF, we deduce
an enthalpy of hydrogenation of  -108.26 kcal/mol, and a calculated
destabilization of 32.72 kcal/mol from the W1 $\Delta H^\circ_{f,298}$
of FCH$_2$--CH$_2$F (-107.83 kcal/mol) and FC$\equiv$CF (0.48 kcal/mol).

Combining the W1 calculated $\Delta H^\circ_{f,298}$[ClCH$_2$CH$_2$Cl]=-33.19
kcal/mol with the W1 $\Delta H^\circ_{f,298}$[ClC$\equiv$CCl]=+56.46 kcal/mol,
we deduce an enthalpy of hydrogenation of -89.60 kcal/mol, and hence a
destabilization of 14.06 kcal/mol. Our calculated heat of formation for
1,2-dichloroethane differs by about 3 kcal/mol 
from the experimental value of -29.97 $\pm$ 0.24 kcal/mol  reported by
Lacher, Amador and Park \cite{500}. This discrepancy is an order of 
magnitude larger than the average error of W1 theory for atomization energies
\cite{w1,w1w2validate}, and the molecule does not exhibit strong nondynamical
correlation effect that could cause failure of the underlying CCSD(T) electron
correlation method. In order to rule out an exceptionally large W1 error,
let us consider enthalpy changes for the following isodesmic reactions:
\begin{eqnarray}
{\rm C}_2{\rm H}_6 + {\rm ClCH}_2{\rm CH}_2{\rm Cl} &\rightarrow& 
2 {\rm ~C}_2{\rm H}_5{\rm Cl} \label{aleph}\\
{\rm C}_2{\rm H}_6 + {\rm CH}_3{\rm Cl} &\rightarrow&
{\rm C}_2{\rm H}_5{\rm Cl} + {\rm CH}_4\label{beit}\\
{\rm C}_2{\rm H}_5{\rm Cl} + {\rm CH}_3{\rm Cl} &\rightarrow& 
{\rm ClCH}_2{\rm CH}_2{\rm Cl} + {\rm CH}_4\label{gimel}\\
{\rm C}_2{\rm H}_6 + 2 {\rm ~CH}_3{\rm Cl} &\rightarrow&
{\rm ClCH}_2{\rm CH}_2{\rm Cl} + 2 {\rm ~CH}_4\label{dalet}
\end{eqnarray}
For reaction (\ref{aleph}), the two experimental heats of formation 
of ethyl chloride, -26.84 and -25.72 kcal/mol, would lead to
$\Delta H_r$=-3.63 and -1.39 kcal/mol, respectively. At the W1 level,
we find $\Delta H_r$=-1.13 kcal/mol, a number that should be basically
exact for a simple isodesmic reaction of the eq. (\ref{aleph}) type.
Hence the first combination of experimental values can be rejected 
out of hand. The second one would imply that W1 theory be 2 kcal/mol
off for ethyl chloride.

The $\Delta H_r$=-3.42 kcal/mol 
for eq. (\ref{gimel}) should likewise be basically
exact. The two experimental value combinations would lead to
$\Delta H_r$=-1.43 and -2.55 kcal/mol, respectively: again the former
value combination can be rejected outright, while the discrepancy of the
latter with our W1 calculations for a reaction as simple as eq. (\ref{gimel})
is somewhat hard to accept. For eq. (\ref{beit}), 
the experimentally derived $\Delta H_r$=-5.06 and -3.94
kcal/mol, respectively, bracket the W1 value, -4.55 kcal/mol, and hence
do not allow us to make a decision. Combining eqs. (\ref{beit},\ref{gimel})
into eq.(\ref{dalet}), however, we can eliminate the ethyl chloride value:
again considering that we are dealing with a simple isodesmic reaction,
the experimentally derived $\Delta H_r$=-6.49 kcal/mol is rather hard
to reconcile with the W1 value of -7.97 kcal/mol. If we combine the W1 isodesmic
reaction energy with the experimental data (and accompanying error bars)
ethane, methane, and methyl
chloride, we obtain a revised $\Delta H^\circ_{f,298}$=-31.45$\pm$0.33
kcal/mol (the errors being treated as standard deviations, not upper
limits).

Alternatively, if we consider the W1 error for ethane to
be transferable to substituted ethanes, then we find 
$\Delta H^\circ_f$[C$_2$H$_5$Cl]=-26.65 kcal/mol, within the error bar
of the Fletcher and Pilcher value. Analogously, we obtain 
$\Delta H^\circ_f$[ClCH$_2$CH$_2$Cl]=-32.09 kcal/mol. Taking the average of
both approaches, and conservatively setting the error bar to twice the
difference between the values, we obtain a best estimate
$\Delta H^\circ_f$[ClCH$_2$CH$_2$Cl]=-31.8$\pm$0.6 kcal/mol.

After the present paper was submitted for publication, we received a
preprint\cite{manion} of a critically evaluated
review of experimental thermochemical data for small chlorocarbons and
chlorohydrocarbons. In said review,  
$\Delta H^\circ_f$[ClCH$_2$CH$_2$Cl]=-133.0$\pm$3.0~kJ/mol,
or -31.8$\pm$0.7~kcal/mol, is proposed, in excellent agreement with our best
theoretical predictions.

Summarizing the destabilization energies of substituted haloacetylenes
(FCCF=32.72, ClCCF=23.60, HCCF=15.58, ClCCCl=14.06 and HCCCl=7.07 kcal/mol),
we can clearly see that the destabilization decreases in the order 
FCCF $>$ ClCCF $>$ HCCF $>$ ClCCCl $>$ HCCCl, in which fluorine and
chlorine atoms contribute about 16 and 7 kcal/mol, respectively.

Finally, it is worthwhile to consider the component breakdown 
of the reaction energies (Table VI) which shows that the
hydrogenation enthalpies to be almost entirely governed by SCF 
contribution. Core correlation contribution is noticable for fluorine 
containing systems while the contribution from scalar relativistic 
effect is less than 0.1 kcal/mol in all cases. The first-order spin-orbit 
contribution trivially cancels for the reaction energies.

An evaluation of the performance of some common lower-cost
thermochemistry methods (such as G2, G3, and CBS-QB3) would appear
to be relevant here. The discrepancies in TAE$_0$ with the benchmark
W2h results are summarized in Table VII. Deviations for W1 theory are
quite moderate, with C$_2$F$_2$ being the only outlier (+0.75 kcal/mol):
with this one exception, individual errors stay well below 0.5 kcal/mol, and
the mean absolute deviation (MAD) is 0.27 kcal/mol, slightly higher than the
average accuracy of W2/W2h theory itself over its training set. The next
best performer is G3 theory (MAD=0.51 kcal/mol), which however has a 
pretty large error of -1.68 kcal/mol for CH$_3$Cl. While we have seen
in the past that CBS-QB3 performs remarkably well on some molecules
that are highly problematic for G3 theory (like SiF$_4$\cite{sif4} and
SO$_3$\cite{so3tae}), its MAD for the present systems is 0.75 kcal/mol,
substantially higher than G3 theory. Four systems (as opposed to a
single one for G3 theory) have errors exceeding 1 kcal/mol. Finally,
the present results do show that G3 theory represents a considerable
improvement over G2 theory (MAD=0.90 kcal/mol). Interestingly, the
more recent G3X method\cite{g3x} in fact performs more poorly
than G3 theory for the problem under study (MAD =0.68 kcal/mol, worst-case
1.36 kcal/mol for C$_2$F$_2$).

\section{Conclusions}

Benchmark ab initio calculations have been carried out on the experimentally
poorly known heats of formation of the fluoroacetylenes XCCY (X,Y=H,F,Cl).
Our best calculated values, obtained using a minor variant of the recently
proposed Weizmann-2 (W2) method, are, at 0 (298) K: 
HCCH 54.66 (54.48) kcal/mol;
HCCF 25.07 (25.15) kcal/mol; FCCF 0.79(1.38) kcal/mol; 
HCCCl 54.69(54.83) kcal/mol; C$_2$Cl$_2$ 55.49(56.21) kcal/mol;
and FCCCl 27.85(28.47) kcal/mol. 
To these values we conservatively affix error bars of 0.50 kcal/mol, i.e. slightly 
more than twice the mean absolute error of W2 theory for a "training set" of 
experimentally very precisely known heats of formation. Analysis of the results 
using isodesmic reactions reveals that some of the assumptions on which the 
experimentally derived estimates are based in fact do not hold particularly well 
in this case.  Calculated hydrogenation energies 
of substituted haloacetylenes suggest that fluorine and chlorine atom 
substitutions destabilize the acetylene by nearly 16 and 7 kcal/mol 
apiece, respectively.
By a combination of W1 theory and isodesmic reactions, we show that
the established heat of formation of 1,2-dichloroethane should be
revised to -31.8$\pm$0.6 kcal/mol, in excellent agreement with a very recent
critically evaluated review.
Comparing our best heat of formation values with those obtained 
using various computationally less expensive schemes, we find that agreement
is best for W1 theory, followed by G3 theory; the G3X and CBS-QB3 method appears to 
perform more poorly for this particular problem. 

\acknowledgments

SP acknowledges a Clore Postdoctoral Fellowship.
JM is the incumbent of the Helen and Milton A. Kimmelman Career 
Development Chair. 
Research at the Weizmann Institute was supported by the Minerva 
Foundation, Munich, Germany. JFL wishes to thank the Dow Chemical 
Company (Midland, Michigan, USA) for partial support of his 
thermochemical studies. This paper
is in honor of Professor Ernest R. Davidson on the 
occasion of his 65th birthday.

\begin{table}
\caption{Reference geometries (\AA, degree) obtained in this work and
used in the thermochemical calculations.}
\squeezetable
\begin{tabular}{lll}
 & B3LYP/cc-pVTZ+1 & CCSD(T)/cc-pVQZ+1 \\
 & for W1 theory & for W2h theory\\
\hline
 & B3LYP/cc-pVTZ+1 & CCSD(T)/cc-pVQZ+1\\
CH$_4$ & r(CH)=1.0885 & r(CH)=1.0879\\
CH$_3$F & r(CF)=1.3865; r(CH)=1.0904; $\theta$(ClCH)=109.06 & r(CF)=1.3824; r(CH)=1.0888; $\theta$(ClCH)=108.91\\
CH$_3$Cl & r(CCl)=1.7959; r(CH)=1.0849; $\theta$(ClCH)=108.34 & r(CCl)=1.7829; r(CH)=1.0851; $\theta$(ClCH)=108.42\\
C$_2$H$_2$ & r(CH)=1.0619; r(CC)=1.1963 & r(CH)=1.0634; r(CC)=1.2065\\
FCCCl & r(CCl)=1.6416; r(CC)=1.1917; r(CF)=1.2802 & r(CCl)=1.6452; r(CC)=1.1975; r(CF)=1.2810\\
HCC$^+$ & r(CH)=1.0940; r(CC)=1.1935 & r(CH)=1.0926; r(CC)=1.2322\\
HCC$^-$ & r(CH)=1.0682; r(CC)=1.2385 & r(CH)=1.0707; r(CC)=1.2493\\
HCCF & r(CH)=1.0600; r(CC)=1.1919; r(CF)=1.2779 & r(CH)=1.0611; r(CC)=1.1995; r(CF)=1.2791\\
C$_2$F$_2$ & r(CC)=1.1838; r(CF)=1.2857 & r(CC)=1.1892; r(CF)=1.2859\\
HCCCl & r(CH)=1.0606; r(CC)=1.1980; r(CCl)=1.6371 & r(CH)=1.0626; r(CC)=1.2063; r(CCl)=1.6412\\
C$_2$Cl$_2$ & r(CC)=1.1996; r(CCl)=1.6365 & r(CC)=1.2055; r(CCl)=1.6408\\
\end{tabular}
\end{table}

\begin{table}
\caption{W1 and W2h total atomization energies (kcal/mol) and 
constituent components thereof; computed and literature heats of formation.
All values in kcal/mol.}
\squeezetable
\begin{tabular}{lddddddddd}
 & CH$_4$ & CH$_3$F & CH$_3$Cl & C$_2$H$_2$ & HCCF & C$_2$F$_2$ & HCCCl & C$_2$Cl$_2$ & FCCCl\\
\hline
W1 theory\\
\hline
TAE[SCF,D] &               325.61 & 313.01 & 297.75 & 290.68 & 264.94 & 234.15 & 256.91 & 221.87 & 228.80  \\
TAE[SCF,T] &               330.96 & 319.16 & 303.17 & 299.47 & 275.89 & 246.92 & 265.97 & 231.37 & 239.81  \\
TAE[SCF,Q] &               331.38 & 319.55 & 303.68 & 300.32 & 276.78 & 247.84 & 267.01 & 232.57 & 240.87  \\
TAE[SCF,$\infty$] &        331.52 & 319.67 & 303.83 & 300.59 & 277.06 & 248.13 & 267.33 & 232.95 & 241.20  \\
$\Delta$TAE[CCSD,D] &      68.67 & 80.02 & 68.16 & 73.02 & 86.06 & 99.20 & 76.04 & 79.32 & 89.13           \\
$\Delta$TAE[CCSD,T] &      79.93 & 90.90 & 79.57 & 86.99 & 99.70 & 112.50 & 89.43 & 92.00 & 102.06         \\
$\Delta$TAE[CCSD,Q] &      83.02 & 94.82 & 83.54 & 91.54 & 104.93 & 118.42 & 94.47 & 97.52 & 107.76        \\
$\Delta$TAE[CCSD,$\infty$] & 85.05 & 97.39 & 86.14 & 94.53 & 108.36 & 122.31 & 97.78 & 101.14 & 111.50     \\
$\Delta$TAE[(T),D] &       1.86 & 3.36 & 3.63 & 5.85 & 7.65 & 9.38 & 8.15 & 10.56 & 9.93                   \\
$\Delta$TAE[(T),T] &       2.64 & 4.86 & 4.84 & 7.59 & 10.11 & 12.53 & 10.32 & 13.12 & 12.77               \\
$\Delta$TAE[(T),$\infty$] & 2.93 & 5.41 & 5.29 & 8.24 & 11.02 & 13.70 & 11.13 & 14.07 & 13.82              \\
Core corr. &               1.21 & 1.12 & 1.19 & 2.40 & 2.60 & 2.82 & 2.61 & 2.83 & 2.82                    \\
Scalar relat. &            -0.19 & -0.37 & -0.42 & -0.27 & -0.48 & -0.70 & -0.55 & -0.84 & -0.77           \\
Spin-orbit &               -0.08 & -0.47 & -0.93 & -0.17 & -0.55 & -0.94 & -1.01 & -1.85 & -1.40           \\
TAE$_e$(W1) &           420.42 & 422.75 & 395.10 & 405.32 & 398.00 & 385.32 & 377.29 & 348.30 & 367.18  \\
ZPVE &                    27.56 & 24.16 & 23.25 & 16.68 & 12.62 & 8.31 & 11.86 & 6.90 & 7.66              \\
TAE$_0$ &                 392.86 & 398.59 & 371.85 & 388.64 & 385.38 & 377.01 & 365.42 & 341.40 & 359.52  \\
$\Delta H^\circ_f$(0 K) & -16.34 & -55.24 & -18.38 & 54.59 & 24.68 & -0.11 & 54.76 & 55.74 & 27.50        \\
$\Delta(H_{298}-H_0)_f$ & -1.91 & -1.92 & -1.89 & -0.17 & 0.08 & 0.59 & 0.14 & 0.72 & 0.62                \\
$\Delta H^\circ_f$(298) & -18.25 & -57.16 & -20.27 & 54.41 & 24.76 & 0.48 & 54.90 & 56.46 & 28.12         \\
\hline
W2h theory\\
\hline
TAE[SCF,T] & 330.97 & 319.18 & 303.19 & 298.66 & 274.97 & 245.91 & 265.19 & 230.74 & 238.98\\
TAE[SCF,Q] & 331.40 & 319.67 & 303.71 & 299.63 & 276.12 & 247.26 & 266.38 & 232.06 & 240.29\\
TAE[SCF,5] & 331.52 & 319.78 & 303.85 & 299.84 & 276.35 & 247.50 & 266.59 & 232.29 & 240.51\\
TAE[SCF,$\infty$] & 331.57 & 319.82 & 303.90 & 299.89 & 276.41 & 247.56 & 266.63 & 232.34 & 240.56\\
$\Delta$TAE[CCSD,T] & 79.46 & 90.40 & 78.78 & 86.93 & 99.48 & 112.11 & 88.86 & 90.86 & 101.33\\
$\Delta$TAE[CCSD,Q] & 82.90 & 94.49 & 83.21 & 91.74 & 104.96 & 118.25 & 94.44 & 97.16 & 107.53\\
$\Delta$TAE[CCSD,5] & 83.83 & 95.66 & 84.59 & 93.26 & 106.63 & 120.06 & 96.32 & 99.42 & 109.55\\
$\Delta$TAE[CCSD,$\infty$] & 84.81 & 96.89 & 86.03 & 94.86 & 108.39 & 121.96 & 98.30 & 101.80 & 111.67\\
$\Delta$TAE[(T),T] & 2.60 & 4.81 & 4.74 & 7.65 & 10.17 & 12.58 & 10.32 & 13.07 & 12.77\\
$\Delta$TAE[(T),Q] & 2.79 & 5.12 & 5.05 & 8.07 & 10.70 & 13.22 & 10.88 & 13.75 & 13.43\\
$\Delta$TAE[(T),$\infty$] & 2.94 & 5.34 & 5.28 & 8.37 & 11.08 & 13.69 & 11.29 & 14.24 & 13.92\\
Core corr. & 1.21 & 1.14 & 1.21 & 2.34 & 2.55 & 2.79 & 2.55 & 2.77 & 2.77\\
Scalar relat. & -0.19 & -0.37 & -0.42 & -0.28 & -0.48 & -0.70 & -0.56 & -0.84 & -0.77\\
Spin-orbit & -0.08 & -0.47 & -0.93 & -0.17 & -0.55 & -0.94 & -1.01 & -1.85 & -1.40\\
TAE$_e$(W2h) & 420.25 & 422.34 & 395.07 & 405.01 & 397.39 & 384.35 & 377.20 & 348.46 & 366.75\\
ZPVE & 27.56 & 24.16 & 23.25 & 16.68 & 12.62 & 8.31 & 11.86 & 6.90 & 7.66\\
better&27.60b& 24.16i& 23.25i& 16.44c& 12.40d& 8.24e& 11.70f& 6.82g& 7.58h\\
TAE$_0$ & 392.65 & 398.18 & 371.82 & 388.57 & 384.99 & 376.11 & 365.50 & 341.64 & 359.17\\ 
$\Delta H^\circ_f$(0 K) & -16.14 & -54.83 & -18.35 & 54.66 & 25.07 & 0.79 & 54.69 & 55.49 & 27.85\\
$\Delta(H_{298}-H_0)_f$ & -1.91 & -1.92 & -1.89 & -0.17 & 0.08 & 0.59 & 0.14 & 0.71 & 0.62\\
$\Delta H^\circ_f$(298) & -18.05 & -56.75 & -20.24 & 54.48 & 25.15 & 1.38 & 54.83 & 56.21 & 28.47\\
\hline
Expt. (a) & -17.889(75) & -56, & -19.59(16), & 54.19(19) & 30 & 5, & 51.1 & 50.1 & ---\\
 & & -59 & -20.53(14) & &  & -45(6) &  &  & \\
\end{tabular}

The notation TAE[SCF,$n$] denotes the total atomization energy at the SCF level with the
appropriate (aug-)cc-pV$n$Z basis set; $\Delta$TAE[CCSD,$n$] denotes the contribution of
the valence correlation energy at the CCSD level; $\Delta$TAE[(T),$n$] that of connected
triple excitations {\em only} at the CCSD(T) level; $n$=$\infty$ denotes results of 
infinite-basis extrapolations (see Methods section).

(a) From the WebBook\cite{webbook}. Uncertainties on last digits indicated in
parentheses.

(b) Accurate ab initio anharmonic force field: T. J. Lee, J. M. L. Martin, and
P. R. Taylor, \JCP{102}{254}{1995}

(c) ditto: Ref.\cite{c2h2}

(d) ditto: Ref.\cite{hccf}

(e) ditto: Ref. \cite{ThielC2F2}.

(f) ZPVE[accurate]-ZPVE[scaled B3LYP] difference interpolated between C$_2$H$_2$ and ClCCCl

(g) Accurate ab initio anharmonic force field: Ref.\cite{ThielC2Cl2}

(h) ZPVE[accurate]-ZPVE[scaled B3LYP] difference interpolated between FCCF and ClCCCl

(i) no accurate anharmonic force field available: B3LYP/cc-pVTZ+1 harmonic ZPVE scaled by 0.985 used instead

\end{table}

\clearpage

\begin{table}
\caption{W2h reaction energies for various isodesmic reactions 
(kcal/mol), as well as constituent components.}
\squeezetable
\begin{tabular}{lcccccccc}
 & C$_2$H$_2$+C$_2$F$_2\rightarrow$ & C$_2$H$_2$+C$_2$Cl$_2\rightarrow$ & FCCF+ClCCCl$\rightarrow$ & C$_2$H$_2$+CH$_3$F$\rightarrow$ & C$_2$H$_2$+CH$_3$Cl$\rightarrow$ & HCCF+CH$_3$F$\rightarrow$ & HCCCl+CH$_3$Cl \\
 & 2 HCCF & 2 HCCCl & 2 FCCCl & HCCF+CH$_4$ & HCCCl+CH$_4$ & C$_2$F$_2$+CH$_4$ & C$_2$Cl$_2$+CH$_4$ \\
\hline
$\Delta E$[SCF,T] & -5.37 & -0.98 & -1.30 & 11.90 & 5.69 & 17.27 & 6.67 \\
$\Delta E$[SCF,Q] & -5.36 & -1.06 & -1.26 & 11.78 & 5.57 & 17.14 & 6.63 \\
$\Delta E$[SCF,5] & -5.36 & -1.05 & -1.23 & 11.75 & 5.58 & 17.11 & 6.62 \\
$\Delta E$[SCF,$\infty$] & -5.36 & -1.03 & -1.22 & 11.73 & 5.58 & 17.09 & 6.62 \\
$\Delta\Delta E$[CCSD,T] & 0.08 & 0.07 & 0.31 & -1.61 & -2.61 & -1.69 & -2.68 \\
$\Delta\Delta E$[CCSD,Q] & 0.08 & 0.02 & 0.34 & -1.62 & -2.38 & -1.71 & -2.40 \\
$\Delta\Delta E$[CCSD,5] & 0.06 & 0.04 & 0.39 & -1.54 & -2.30 & -1.60 & -2.34 \\
$\Delta E$[CCSD,$\infty$] & 0.03 & 0.06 & 0.43 & -1.46 & -2.22 & -1.49 & -2.28 \\
$\Delta\Delta E$[(T),T] & -0.11 & 0.07 & 0.11 & -0.31 & -0.53 & -0.20 & -0.61 \\
$\Delta\Delta E$[(T),Q] & -0.11 & 0.05 & 0.10 & -0.31 & -0.56 & -0.20 & -0.61 \\
$\Delta\Delta E$[(T),$\infty$] & -0.11 & 0.03 & 0.09 & -0.31 & -0.58 & -0.20 & -0.61 \\
$\Delta E$core & 0.03 & 0.01 & 0.02 & -0.28 & -0.21 & -0.32 & -0.22 \\
$\Delta E$(DMV) & -0.01 & 0.00 & 0.00 & 0.03 & 0.05 & 0.04 & 0.05 \\
$\Delta E$(total) & -5.42 & -0.92 & -0.69 & 9.71 & 2.63 & 15.13 & 3.55 \\
ditto W1$^a$  & -5.37 & -0.95 & -0.74 & 9.65 & 2.72 & 15.02 & 3.67 \\
$\Delta$ZPVE$^b$ & -0.25 & -0.14 & -0.11 & 0.66 & 0.50 & 0.91 & 0.65 \\
$\Delta H_0$ & -5.18 & -0.78 & -0.58 & 9.05 & 2.12 & 14.22 & 2.90 \\
$\Delta(H_{298}-H_0)$ & -0.25 & -0.25 & -0.06 & 0.27 & 0.30 & 0.52 & 0.56\\
$\Delta H_{298}$ & -5.42 & -1.03 & -0.64 & 9.32 & 2.43 & 14.74 & 3.46 \\
\end{tabular}

See previous table for notation details.

(a) for comparison

(b) scaled harmonic B3LYP/cc-pVTZ+1 ZPVE used throughout for consistency

\end{table}

\begin{table}
\caption{B3LYP/cc-pVTZ+1//CCSD(T)/c-pVQZ+1 Wiberg bond indices derived from
a natural population analysis.}
\squeezetable
\begin{tabular}{ll}
 & NPA Wiberg bond orders $\geq$0.02\\
\hline
CH$_4$ & BO(CH)=0.96\\
CH$_3$F & BO(CH)=0.96; BO(CF)=0.87; BO(HF)=0.03\\
CH$_3$Cl & BO(CH)=0.94; BO(CCl)=1.03; BO(HCl)=0.02\\
C$_2$H$_2$ & BO(CH)=0.94; BO(CC)=3.00\\
HCCF & BO(CF)=1.00; BO(CC)=2.83; BO(CH)=0.93; BO(FC')=0.16\\
C$_2$F$_2$ & BO(CF)=0.98; BO(CC)=2.71; BO(FC')=0.15\\
HCCCl & BO(CCl)=1.19; BO(CC)=2.80; BO(CH)=0.93; BO(ClC')=0.21\\
C$_2$Cl$_2$ & BO(CCl)=1.18; BO(CC)=2.63; BO(ClC')=0.19; BO(ClCl)=0.03\\
FCCCl & BO(CF)=1.00; BO(CC)=2.67; BO(CCl)=1.16; BO(FC')=0.15; BO(ClC')=0.19\\
\end{tabular}
\end{table}

\begin{table}
\caption{Components of W1 total atomization energies (kcal/mol)  
of  ethane and haloethanes.}
\squeezetable
\begin{tabular}{ldddddd}
 & C$_2$H$_6$ & C$_2$H$_5$F & C$_2$H$_5$Cl & FC$_2$H$_4$F & FC$_2$H$_4$Cl & ClC$_2$H$_4$Cl \\
\hline
W1 theory\\
\hline
TAE[SCF,D] &                   548.64  &  541.32  &  523.96  &  530.57  &  514.20  &  497.27     \\
TAE[SCF,T] &                   556.96  &  550.38  &  532.15  &  540.49  &  523.23  &  505.40     \\
TAE[SCF,Q] &                   557.76  &  551.07  &  533.00  &  541.08  &  523.99  &  506.31     \\
TAE[SCF,$\infty$] &            558.01  &  551.29  &  533.26  &  541.26  &  524.23  &  506.59     \\
$\Delta$TAE[CCSD,D] &          117.35  &  129.08  &  117.85  &  140.61  &  129.84  &  119.09     \\
$\Delta$TAE[CCSD,T] &          137.37  &  148.86  &  138.04  &  160.11  &  149.73  &  139.28     \\
$\Delta$TAE[CCSD,Q] &          143.13  &  155.41  &  144.55  &  167.38  &  156.96  &  146.40     \\
$\Delta$TAE[CCSD,$\infty$]  &  146.90  &  159.70  &  148.82  &  172.15  &  161.70  &  151.08     \\
$\Delta$TAE[(T),D] &           4.20  &  5.88  &  6.31  &  7.52  &  8.01  &  8.53                 \\
$\Delta$TAE[(T),T] &           5.82  &  8.21  &  8.36  &  10.54  &  10.76  &  11.01              \\
$\Delta$TAE[(T),$\infty$] &    6.43  &  9.08  &  9.13  &  11.66  &  11.78  &  11.92              \\
Core corr. &                   2.34  &  2.30  &  2.35  &  2.23  &  2.61$^a$  &  2.67$^a$         \\
Scalar relat. &                -0.39  &  -0.55  &  -0.60  &  -0.72  &  -0.76  &  -0.79           \\
Spin-orbit &                   -0.17  &  -0.55  &  -1.01  &  -0.94  &  -1.40  &  -1.85           \\
TAE$_e$(W1) &                  713.12  &  721.25  &  691.95  &  725.63  &  698.16  &  669.62     \\
ZPVE &                         45.97  &  41.85  &  41.01  &  37.63  &  36.73  &  35.81           \\
TAE$_0$ &                      667.15  &  679.40  &  650.94  &  688.00  &  661.43  &  633.81     \\
$\Delta H^\circ_f$(0 K) &      -17.39  &  -62.81  &  -24.23  &  -104.57  &  -67.88  &  -30.13    \\
$\Delta(H_{298}-H_0)_f$ &      -3.79  &  -3.60  &  -3.52  &  -3.26  &  -3.18  &  -3.06           \\
$\Delta H^\circ_f$(298) &      -21.18  &  -66.41  &  -27.75  &  -107.83  &  -71.06  &  -33.19    \\
\end{tabular}
The notation TAE[SCF,$n$] denotes the total atomization energy at the SCF level with the
appropriate (aug-)cc-pV$n$Z basis set; $\Delta$TAE[CCSD,$n$] denotes the contribution of
the valence correlation energy at the CCSD level; $\Delta$TAE[(T),$n$] that of connected
triple excitations {\em only} at the CCSD(T) level; $n$=$\infty$ denotes results of 
infinite-basis extrapolations (see Methods section). \\
(a) Core-correlation contribution is derived using MSFT empirical model\cite{msft}
\end{table}

\clearpage

\begin{table}
\caption{Reaction energies for hydrogenation reactions,
X-C$\equiv$C-Y + 2H$_2$ $\rightarrow$ X-CH$_2$CH$_2$-Y,
calculated at W1 level of theory. All values in kcal/mol.}
\squeezetable
\begin{tabular}{ldddddd}
   & HCCH & HCCF & HCCCl & FCCF & FCCCl & ClCCCl \\
\hline
TAE[SCF,D] &                    94.59  &  113.01  &  103.69  &  133.05  &  122.03  &  112.03     \\
TAE[SCF,T] &                    90.18  &  107.18  &  98.86  &  126.25  &  116.11  &  106.72      \\
TAE[SCF,Q] &                    89.84  &  106.69  &  98.39  &  125.64  &  115.52  &  106.14      \\
TAE[SCF,$\infty$] &             89.73  &  106.54  &  98.24  &  125.44  &  115.34  &  105.96      \\
$\Delta$TAE[CCSD,D] &           0.74  &  -0.56  &  -1.77  &  -2.17  &  -2.87  &  -3.82           \\
$\Delta$TAE[CCSD,T] &           0.93  &  -0.29  &  -0.83  &  -1.83  &  -1.78  &  -2.17           \\
$\Delta$TAE[CCSD,Q] &           0.93  &  -0.18  &  -0.57  &  -1.69  &  -1.46  &  -1.77           \\
$\Delta$TAE[CCSD,$\infty$]  &   0.92  &  -0.10  &  -0.40  &  -1.61  &  -1.25  &  -1.51           \\
$\Delta$TAE[(T),D] &            -1.65  &  -1.77  &  -1.85  &  -1.85  &  -1.92  &  -2.02          \\
$\Delta$TAE[(T),T] &            -1.77  &  -1.90  &  -1.96  &  -1.99  &  -2.01  &  -2.11          \\
$\Delta$TAE[(T),$\infty$] &     -1.81  &  -1.94  &  -2.00  &  -2.04  &  -2.04  &  -2.15          \\
Core corr. &                    -0.07  &  -0.30  &  -0.26  &  -0.60  &  -0.21  &  -0.16          \\
Scalar relat. &                 -0.12  &  -0.07  &  -0.04  &  -0.02  &  0.01  &  0.05            \\
TAE$_e$(W1) &                   88.66  &  104.12  &  95.53  &  121.19  &  111.85  &  102.19      \\
ZPVE &                          16.84  &  16.78  &  16.69  &  16.87  &  16.62  &  16.46          \\
TAE$_0$ &                       71.82  &  87.34  &  78.84  &  104.32  &  95.23  &  85.73         \\
$\Delta H^\circ_f$(0 K) &       -71.82  &  -87.34  &  -78.84  &  -104.32  &  -95.23  &  -85.73   \\
$\Delta(H_{298}-H_0)_f$ &       -3.71  &  -3.79  &  -3.77  &  -3.95  &  -3.90  &  -3.87          \\
$\Delta H^\circ_f$(298) &       -75.54  &  -91.12  &  -82.61  &  -108.26  &  -99.14  &  -89.60   \\
\hline                                                                                                
Destabilization Energy            & &  15.58  &  7.07  &  32.72  &  23.60  &  14.06                \\
\end{tabular}
\end{table}

\clearpage

\begin{table}
\caption{Deviations (kcal/mol) of computed TAE$_0$ values from the 
W2h results.}
\squeezetable
\begin{tabular}{ldddddddddd}
             & W1 & G2 & G3$^{a}$ & G3X & CBS-QB3$^{a}$\\
\hline
CH$_4$       & 0.21 & 0.48 & 0.01  & -0.13   & -0.37\\
CH$_3$F      & 0.41 & 1.61 & -0.36  & -0.28   & 0.32\\
CH$_3$Cl     & 0.03 & 0.23 & -1.68  & -1.00   & -0.25\\
C$_2$H$_2$   & 0.07 & -1.08 & -0.62 &  0.03  & -1.33\\
HCCF         & 0.39 & 0.24 & -0.27  &  0.46  & -0.04\\
C$_2$F$_2$   & 0.90 & 1.52 & 0.32   &  1.36  & 1.36\\
HCCCl        & -0.08 & 0.08 & -0.68 &  0.51  & -0.01\\
C$_2$Cl$_2$  & -0.24 & 1.40 & -0.60 &  1.07  & 1.48\\
FCCCl        & 0.35 & 1.49 & 0.03   &  1.31  & 1.59\\
mean abs.dev.& 0.30 & 0.90 & 0.51   &  0.68  & 0.75\\
\end{tabular}

(a) both G3 and CBS-QB3 numbers include atomic spin-orbit corrections

\end{table}


\begin{thebibliography}{99}
\bibitem{JL3} This is perhaps an understatement. Ref. \cite[b]{JL2} 
cites numerous references mentioning explosive and pyrophoric behavior
\end{thebibliography}
\end{document}